\documentclass[prl,twocolumn,showpacs,preprintnumbers,amsmath,amssymb,superscriptaddress]{revtex4}

\usepackage{graphicx}% Include figure files
\usepackage{dcolumn}% Align table columns on decimal point
\usepackage{bm}% bold math

\def \ed {\end{document}}
\def\Fbox#1{\vskip1ex\hbox to 8.5cm{\hfil\fboxsep0.3cm\fbox{%
  \parbox{8.0cm}{#1}}\hfil}\vskip1ex\noindent}  %%  {TEXT} in BOX
\def\be{\begin{equation}}\def\ee{\end{equation}}
\def\bea{\begin{eqnarray}}\def\eea{\end{eqnarray}}
\def\bse{\begin{subequations}}\def\ese{\end{subequations}}
\newcommand{\BE}[1]{\begin{equation}\label{#1}}
\newcommand{\BEA}[1]{\begin{eqnarray}\label{#1}}
\newcommand{\BSE}[1]{\begin{subequations}\label{#1}}

\let \= \equiv \let\*\cdot \let\~\widetilde \let\^\widehat \let\-\overline

\def\<{\left\langle}    \def\>{\right\rangle}
\def\({\left(}          \def\){\right)}
 \def \[ {\left [} \def \] {\right ]}

\makeatletter
\AtBeginDocument{\@ifpackageloaded{natbib}{\ifNAT@numbers\if@filesw\immediate\write\@auxout{\string\global\string\NAT@numberstrue}\fi\fi}{}}
\makeatother

\begin{document}

\preprint{APS/123-QED}

\title{Spin-superflow turbulence in spin-1 ferromagnetic spinor Bose-Einstein condensates}

\author{Kazuya Fujimoto}
\affiliation{Department of Physics, Osaka City University, Sumiyoshi-ku, Osaka 558-8585, Japan}

\author{Makoto Tsubota}
\affiliation{Department of Physics, Osaka City University, Sumiyoshi-ku, Osaka 558-8585, Japan}
\affiliation{The OCU Advanced Research Institute for Natural Science and Technology (OCARINA), Osaka City University, Sumiyoshi-ku, Osaka 558-8585, Japan}

\date{\today}

\begin{abstract}
Spin-superflow turbulence (SST) in spin-1 ferromagnetic spinor Bose-Einstein condensates is theoretically and numerically studied
by using the spin-1 spinor Gross-Pitaevskii (GP) equations. SST is  turbulence in which the disturbed spin and superfluid velocity fields are coupled.
Applying the Kolmogorov-type dimensional scaling analysis to the hydrodynamic equations of spin and velocity fields,
we theoretically find that the $-5/3$ and $-7/3$ power laws can appear in spectra of the superflow kinetic and the spin-dependent interaction energy, respectively. Our numerical calculation of the GP equations with a phenomenological small-scale energy dissipation confirms SST with the coexistence of disturbed spin and superfluid velocity field with two power laws. 
\end{abstract}


\maketitle

\section{Introduction}
Turbulence is a strong nonequilibrium phenomenon, exhibiting unpredictable behavior of the velocity field, which can result from
the multiple degrees of freedom and  nonlinearity of fluid systems \cite{Davidson,Frisch}.
This situation complicates our understanding of  turbulence, making it one of the unresolved problems in modern physics.

The studies on turbulence in quantum fluids have possibility to shed new light on our understanding of turbulence. 
In quantum turbulence (QT) for quantum fluids such as superfluid helium and one-component atomic Bose-Einstein condensate (BEC) 
\cite{Hal,Vinen,Skrbek12,Nore,Parker,KT05,KT07,Gou09,Henn09,White10,Boris12,Reeves12,Wilson}, 
many quantized vortices with quantized circulation are nucleated, forming tangles. This situation is drastically different 
from classical turbulence (CT) realized in classical fluids because the circulation of velocity field is not quantized in this system. 
Therefore, the element of QT is more obvious than that in CT, and QT is believed to be useful for the understanding of CT.

Recently, a new trend begins to appear in turbulence study in atomic BECs, which is turbulence in multi-component BECs \cite{taketur,FT12a,FT13a,Karl13a,Karl13b,Koba14,Vill14}. 
Atomic BECs have various characteristic features, one of which is a realization of multicomponent BECs \cite{PR,KU,stamper}.
In this system, there exists not only a velocity field but also a (quasi) spin field. 
Owing to the spin degrees of freedom, various topological excitations such as  monopole, skyrmion, knot, domain wall, and 
vortex appear. Therefore, in this system, one expects novel turbulence, 
in which both the velocity and spin fields are much disturbed and various topological excitations are generated.  
Thus, this kind of turbulence should give us new observations for turbulence not found in conventional systems. 

Previously, we studied  turbulence in a spin-1 ferromagnetic spinor BEC, which is a typical multicomponent BEC \cite{FT12a}.
In this turbulence, the spin field is disturbed, so that we call it spin turbulence (ST).
In our previous study, we focused on the spectrum of the spin-dependent interaction energy corresponding to
the spin correlation, finding the characteristic $-7/3$ power law. However, this observation sees only one side of ST. In ST, the velocity and spin fields interact 
with each other, so that a coupled turbulence with two fields can be realized, lending the possibility of showing a property of the velocity field 
not seen in conventional CT and QT. Studying the both sides of turbulence completes the story.  

In this paper, we treat this problem, focusing on the spectrum of the superflow kinetic energy in a spin-1 ferromagnetic spinor BEC.
The spectrum of the kinetic energy is theoretically and numerically found to show a Kolmogorov spectrum
through the interaction between the spin and velocity fields.
The Kolmogorov spectrum refers to the $-5/3$ power law in the kinetic energy spectrum, which is known to appear in CT and QT \cite{Davidson,Frisch,Hal}.
This spectrum is considered to be related to the vortex dynamics, so that it is significant to confirm it
for understanding the turbulence addressed here.
Furthermore, when the $-5/3$ power law is sustained, the spectrum of the spin-dependent interaction energy exhibits a $-7/3$ power law simultaneously.
Therefore, we obtain the coupled turbulence with the disturbed spin and superfluid velocity fields
sustaining the two power laws, anew calling it spin-superflow turbulence (SST) instead of ST.

\section{Formulation}
We consider a BEC of spin-1 bosonic atoms with mass $M$ at zero temperature without trapping and magnetic fields. This system is well described by
the macroscopic wave functions $\psi _m$ ($m = 1,0,-1$) with  magnetic quantum number $m$, which
obey the spin-1 spinor Gorss-Pitaevskii (GP) equations \cite{Ohmi98, Ho98} given by
\begin{eqnarray}
i\hbar \frac{\partial}{\partial t} \psi _{m}  &=&  -\frac{\hbar ^2 }{2M} \bm{\nabla} ^2   \psi _{m} + c_{0} \rho \psi _{m} + c_{1}  \bm{F} \cdot {\hat{\bm{F}} } _{mn} \psi _{n}.   \label{GP}
\end{eqnarray}
In this paper, Roman indices that appear twice are to be summed over $-1$, 0, and 1, and Greek indices are to be summed over $x$, $y$, and $z$.
The parameters $c_{0}$ and $c_{1}$ are the coefficients of the spin-independent and spin-dependent interactions, which are expressed by
$4 \pi \hbar ^{2} (a_{0} + 2a_{2})/3M$ and $4 \pi \hbar ^{2} (a_{2}  - a_{0})/3M$, respectively.
Here, $a_{0}$ and $a_{2}$ are the $s$-wave scattering lengths corresponding to the total spin-$0$ and spin-2 channels.
The total density $\rho$ and the spin density vector $F_{\mu}$ ($\mu = x, y, z$ ) are given by $\rho =  \psi _m ^{*} \psi _m $ and  $F_{\mu} =  \psi _{m}^{*} ({\hat F}_{\mu})_{mn} \psi _{n}$, where $({\hat F}_{\mu})_{mn}$ are the spin-1 matrices.
The sign of the coefficient $c_{1}$ drastically changes the spin dynamics. In this paper, we consider the ferromagnetic interaction $c_{1} < 0$.

In this paper, we focus on the energy spectra for the superflow kinetic and the spin-dependent interaction energy.
The kinetic energy of superfluid velocity $\bm{v}$ per unit mass is given by
\begin{eqnarray}
E_{\rm v} = \frac{1}{2N} \int d\bm{r}  \bm{A}_{\rm v}(\bm{r})^{2},
\end{eqnarray}
where $\bm{A}_{\rm v} = \sqrt{\rho} \bm{v}$ and $N$ is a total particle number.
The superfluid velocity $v_{\mu}$ is given by
\begin{eqnarray}
v_{\mu} = \frac{\hbar}{2M \rho i} \biggl (    \psi _{m}^{*} \nabla_{\mu}  \psi _{m} -  \psi _{m} \nabla_{\mu}  \psi _{m}^{*}  \biggr ).
\label{velocity}
\end{eqnarray}
By using the Fourier series $\bm{A}_{\rm v} (\bm{r}) = \sum _{\bm{k}} \tilde {\bm{A}}_{\rm v} (\bm{k}) e^{i \bm{k} \cdot \bm{r}}$,
we define the spectrum for the kinetic energy per unit mass:
\begin{eqnarray}
\mathcal{E}_{\rm v}(k) = \frac{1 }{2 \rho _{0} \Delta k}  \sum _{k<|\bm{k}_{1}|<k+\Delta k}    | \tilde{\bm{A}}_{\rm v}(\bm{k}_{1})|^{2},
\end{eqnarray}
where $\Delta k$ and $\rho _{0}$ are given by $2 \pi /L$ and $N/{L}^{n_{\rm d}}$, respectively, with  system size $L$ and  spatial dimension $n_{\rm d}$.
Similarly, the spectrum of the spin-dependent interaction energy per unit mass is defined by
\begin{eqnarray}
\mathcal{E}_{\rm s}(k) = \frac{c_{1} }{2 M \rho_{0} \Delta k}  \sum _{k<|\bm{k}_{1}|<k+\Delta k}   | \tilde{\bm{F}}(\bm{k}_{1})|^{2},
\end{eqnarray}
where $ \tilde{\bm{F}}(\bm{k})$ is defined by $\mathcal{F}\large{[} \bm{F}(\bm{r}) \large{]} $ with the Fourier transformation 
$\mathcal {F} \large{[ \cdot]} = \int \cdot \hspace{0.5mm}e^{-i\bm{k}\cdot \bm{r}}  d\bm{r}/V$ and $V = L^{n_{\rm d}}$.

\section{Kolmogorov spectrum and spin-1 spinor GP equations}
We discuss the possibility for a Kolmogorov spectrum in SST with hydrodynamic equations obtained from Eq. (\ref{GP}).  
These hydrodynamic equations are derived in \cite{YU12}, being composed of the equations of the total density $\rho$, 
the superfluid velocity $\bm{v}$, 
the spin vector $f_{\mu} = F_{\mu}/\rho$, and the nematic tensor $n_{\mu \nu} = \psi _{m}^{*} (\hat{N}_{\mu \nu})_{mn} \psi _{n} /\rho$ 
with $(\hat{N}_{\mu \nu})_{mn}  =  [ (\hat{F}_{\mu})_{ml} (\hat{F}_{\nu})_{ln} + (\hat{F}_{\nu})_{ml} (\hat{F}_{\mu})_{ln} ]/2$.

We apply the following three approximations to these hydrodynamic equations: (i) the macroscopic wave functions are expressed by the fully magnetized state, (ii) the total density is almost uniform ($\rho (t) \sim \rho _{0}$), and (iii) the magnitude of velocity $\bm{v}$ is much smaller than the density sound velocity $C_{\rm d} = \sqrt{c_{0}\rho_{0}/2M}$. These similar approximations are discussed in our previous papers \cite{FT12a, FT13a}.
The approximation (i) leads to the relation between the spin vector and nematic tensor \cite{YU12}:
\begin{eqnarray}
n_{\mu\nu} = \frac{\delta _{\mu \nu} + f_{\mu} f_{\nu}}{2}.
\end{eqnarray}
Thus, by eliminating the nematic tensor in the hydrodynamic equations of \cite{YU12}, we obtain the following equations 
\begin{eqnarray}
\frac{\partial}{\partial t} \rho f_{\mu} +  \bm{\nabla} \cdot \rho \bm{v}_{\mu} = 0, \label{1}
\end{eqnarray}
\begin{eqnarray}
\bm{v}_{\mu} = f_{\mu} \bm{v} - \frac{\hbar}{2M} \epsilon _{\mu \nu \lambda}  f_{\nu} (\bm{\nabla} f_{\lambda}), \label{2}
\end{eqnarray}
\begin{eqnarray}
\frac{\partial}{\partial t}  v_{\mu} &+& v_{\nu} \nabla _{\nu} v_{\mu} - \frac{\hbar ^{2}}{2M^{2}} \nabla _{\mu} \frac{\nabla ^{2}_{\nu} \sqrt{\rho}}{\rho}  \nonumber  \\ 
&+& \frac{\hbar^{2}}{4M^{2}\rho} \nabla _{\nu} \rho \biggl\{ (\nabla_{\mu}f_{\lambda})(\nabla_{\nu}f_{\lambda}) 
- f_{\lambda}(\nabla_{\mu} \nabla_{\nu} f_{\lambda})  \biggr\}  \nonumber  \\ 
&=& -\frac{1}{M}\biggl\{ c_{0}(\nabla _{\mu} \rho) + c_{1} f_{\nu} (\nabla _{\mu} \rho f_{\nu}) \biggr\} . \label{3}
\end{eqnarray}
With the approximations (ii) and (iii), Eqs. (\ref{1}) - (\ref{3}) become 
\begin{eqnarray}
\frac{\partial}{\partial t}  f_{\mu}  \sim \frac{\hbar}{2M} \epsilon _{\mu \nu \lambda}  \bm{\nabla} \cdot  [ f_{\nu} (\bm{\nabla} f_{\lambda})],
\label{spin_motion}
\end{eqnarray}
\begin{eqnarray}
\frac{\partial}{\partial t}  v_{\mu} + v_{\nu}\nabla _{\nu}v_{\mu}  \sim  - \frac{\hbar^{2}}{4M^{2}} \nabla _{\nu}  \biggl\{ (\nabla_{\mu}f_{\lambda})(\nabla_{\nu}f_{\lambda})  \nonumber \\
-f_{\lambda}(\nabla_{\mu} \nabla_{\nu} f_{\lambda})  \biggr\}.  
\label{velocity_motion}
\end{eqnarray}
In Eq. (\ref{velocity_motion}), the inertial term $v_{\nu}\nabla _{\nu}v_{\mu}$ is smaller than the other terms \cite{oe},
but we retain this term for the following explanation.

We apply a Kolmogorov-type dimensional scaling analysis \cite{scaling1,scaling2} to Eqs. (\ref{spin_motion}) and
(\ref{velocity_motion}), obtaining the $-5/3$ power law.
We consider the scale transformation $\bm{r}\rightarrow \alpha \bm{r}$ and $ t \rightarrow \beta t$.
Then, if $f_{\mu}$ and $v _{\mu}$ are transformed to $f_{\mu} \rightarrow \alpha^{2}\beta^{-1} f_{\mu}$
and $v_{\mu} \rightarrow \alpha \beta^{-1} v_{\mu}$,
Eqs. (\ref{spin_motion}) and (\ref{velocity_motion}) are invariant.
Thus, the velocity field satisfies $v_{\mu} \sim \Lambda _{\rm v} rt^{-1}$ with a nondimensional coefficient $\Lambda _{ \rm v}$.
Then the spatial and temporal dependence of $\bm{A}_{ \rm v}$ is same as that of $\bm{v}$, which leads to $({\bm{A}}_{ \rm v})_{\mu} \sim \Lambda _{\rm  A} rt^{-1}$ with $\Lambda _{\rm  A} = \sqrt{\rho_{0}} \Lambda _{\rm v}$.
Thus, in SST, the spectrum of the superflow kinetic energy can
be determined by the kinetic energy flux $\epsilon _{\rm v}$ and the coefficient $\Lambda _{\rm v}$,
which, by using  a Kolmogorov-type dimensional analysis, leads to
\begin{eqnarray}
\mathcal{E}_{\rm v}(k) \sim \Lambda _{\rm v}^{2/3} \epsilon _{\rm  v}^{2/3}k^{-5/3}.
\end{eqnarray}
The coefficient $\Lambda _{\rm v}^{2/3}$ is nondimensional, which corresponds to the Kolmogorov constant in CT.
Therefore, the spectrum of the kinetic energy of SST can obey the Kolmogorov spectrum.

Applying a similar analysis to the spin field, we obtain the relation $f_{\mu} \sim \Lambda _{\rm  f} r^{2}t^{-1}$ with 
a dimensional coefficient $\Lambda _{\rm  f}$, which 
leads to the $-7/3$ power law given by
\begin{eqnarray}
\mathcal{E}_{\rm s}(k) \sim \Lambda _{\rm  s}^{2/3} \epsilon _{\rm s}^{2/3}k^{-7/3}
\end{eqnarray}
with the spin-dependent interaction energy flux $\epsilon _{\rm s}$ and a dimensional coefficient
$\Lambda _{\rm  s} = \Lambda _{\rm  f} \sqrt{ |c_{1}|\rho _{0}/M }$.
This was discussed in the previous study \cite{FT12a,FT13a}.

We note that this $-5/3$ power law in SST is much different from that in CT. 
In three (two)-dimensional CT, there is the direct (inverse) energy cascade, where the $-5/3$ power law is generated by
the inertial term $v_{\nu}\nabla _{\nu}v_{\mu}$ in the Navier-Stokes equation \cite{Davidson,Frisch}.  
On the contrary, in SST, the spatial gradient of the spin vector in Eq. (\ref{velocity_motion}) leads to the $-5/3$ power law because, 
in Eq. (\ref{velocity_motion}), the order estimation \cite{oe} finds that the inertial term is smaller than the nonlinear spin term in the range 
$k\xi_{\rho} \agt 0.1$ with the density coherence length $\xi _{\rho} = \hbar / \sqrt{2Mc_{0} \rho _{0}}$.
This suggests that the mechanism responsible for the $-5/3$ power law in SST should be different from that in CT.

Finally, the assumptions used in the derivation of the $-5/3$ and $-7/3$ power laws are discussed. 
We use five assumpotions, which are (i) the macroscopic wave functions are expressed by the fully magnetized state, 
(ii) the total density is almost uniform ($\rho (t) \sim \rho _{0}$), (iii) the magnitude of velocity $\bm{v}$ is much smaller than 
the density sound velocity $C_{\rm d}$, (iv) the spin and velocity fields are scale invariant, and (v) the energy flux is 
independent of the wave number. It is difficult to theoretically confim the validity of these assumptions.  
Then, we consider that the validity may be indirectly confirmed if the $-5/3$ and $-7/3$ power laws based on these assumptions 
appear in the numerical result. 

In the following, we show our numerical method and results to confirm these theoretical considerations.

\section{Numerical method}

\begin{figure} [b]
\begin{center}
\includegraphics[width=85mm]{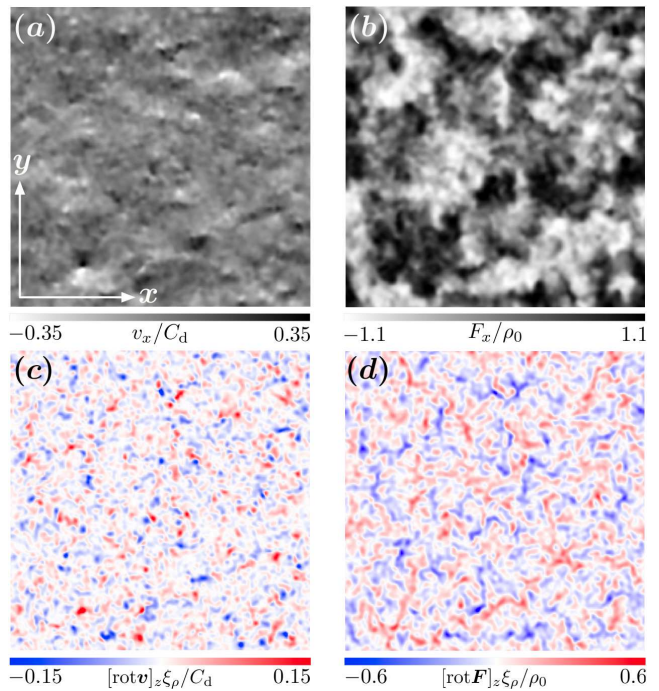}
\caption{(Color online) Spatial distribution of the $x$ components and the rotations for the velocity and spin fields in SST at $t /
\tau = 700$.
(a) and (b) show the $x$ components for the velocity $\bm{v}$ and spin $\bm{F}$ fields, and (c) and (d) show the $z$ component of their rotation.
The system size is $256\xi _{\rho} \times 256\xi _{\rho}$.}
\label{fig1}
\end{center}
\end{figure}

\begin{figure*} [t]
\begin{center}
\includegraphics[width=180mm]{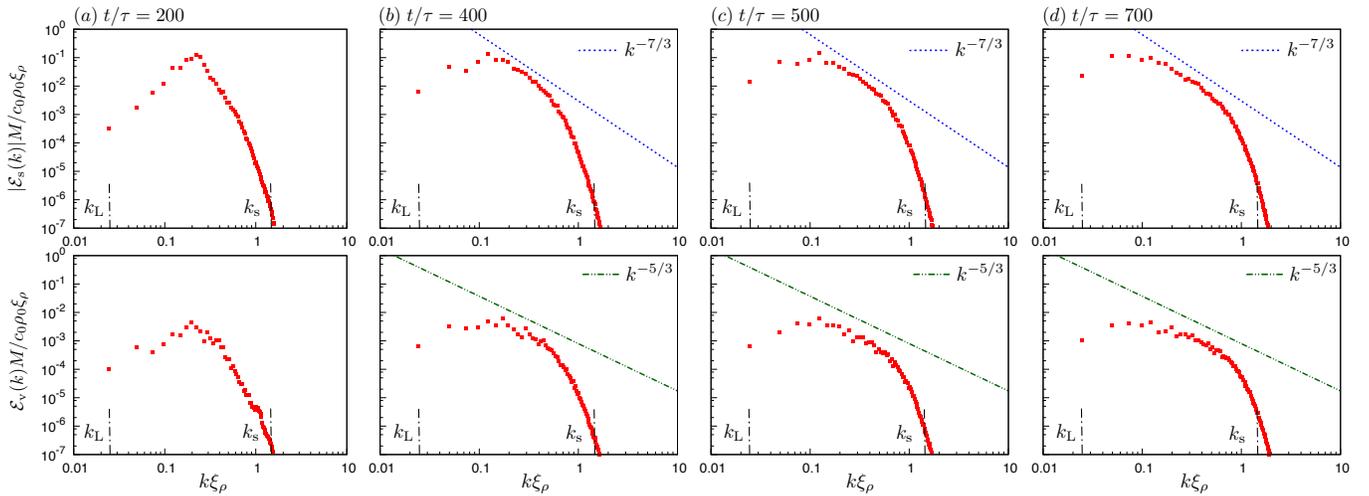}
\caption{(Color online) Time dependence of the spectrum of spin-dependent interaction (upper) and  superflow kinetic (lower) energy at
$t/\tau =$ (a)  200, (b) 400, (c) 500, and (d) 700. The dotted lines in the upper and lower graphs are proportional to $k^{-7/3}$ and $k^{-5/3}$, respectively.}
\label{fig2}
\end{center}
\end{figure*}

\subsection{Small-scale energy dissipation}

In our numerical calculation, we introduce a phenomenological small-scale energy dissipation term into Eq. (\ref{GP}) for the following reasons.
In three-dimensional CT, an energy cascade from low to high wave number is assumed, and
the energy in the high-wave-number region is considered to dissipate \cite{Davidson,Frisch}.
Unless this kind of dissipation takes place,  energy accumulates in the high-wave-number region, 
which can break the power law in the spectrum.

Based on this description of the energy cascade in three-dimensional CT, in the previous study of QT in the one-component BECs \cite{KT05}, 
a phenomenological small-scale dissipation was added to the one-component GP equation given by
\begin{eqnarray}
\bigl( i-\gamma (\bm{k}) \bigr) \hbar \frac{\partial}{\partial t}  \tilde{\psi} (\bm{k})  &=& \frac{\hbar ^2 k^{2} }{2M} \tilde{\psi} (\bm{k})  \nonumber \\
&-& \mu (t) \tilde{\psi} (\bm{k}) + p(\bm{k}),
\label{d1GPk}
\end{eqnarray}
\begin{eqnarray}
p(\bm{k}) = \mathcal{F} \large{[} g |\psi (\bm{r})|^{2}\psi (\bm{r}) \large{]},
\end{eqnarray}
where the macroscopic wave function and the interaction coefficient for the one-component BEC
are denoted by ${\psi}$ and $g$, respectively. $\tilde{\psi} (\bm{k})$ is the Fourier component $\mathcal{F}\large{[} \psi (\bm{r}) \large{]} $ of 
this wave function. 
The function $\gamma (\bm{k})$ is defined by $\gamma _{0} \theta (k-k_{\rho}^{\rm one})$ with the step function $\theta$ and the strength of dissipation $\gamma _{0}$, which
dissipates the energy in the high-wave-number region larger than the wave number $k_{\rho}^{\rm one}$ corresponding to the coherence length 
in the one-component GP equation.
Also, this dissipation reduces the particle number, so that, in the study of Ref. \cite{KT05}, the chemical potential is adjusted for it's conservation.
Thus the chemical potential $\mu(t)$ has a time dependence.

We introduce a similar phenomenological dissipation term into the spin-1 spinor GP equations (\ref{GP}). 
In a spin-1 spinor BEC, there are two characteristic lengths:  the density coherence length $\xi _{\rho}$ and the
spin coherence length  $\xi _{\rm s} $, which are defined by $\hbar / \sqrt{2Mc_{0} \rho _{0}}$ and $\hbar/ \sqrt{2M |c_{1}| \rho _{0}}$. 
In the usual experiments, $|c_{0}/c_{1}|$ is larger than unity, which leads to the condition $\xi _{\rho} < \xi _{\rm s}$.
The size of the spin structure such as the spin domain wall and the spin vortex is on the order of $\xi_{\rm s}$. 
With reference to QT in the one-component BEC, 
we expect that the energy dissipation occurs for wave numbers greater than the wave number $k_{\rm s}$
corresponding to the spin coherence length $\xi _{\rm s}$.
Therefore, we add a phenomenological small-scale energy dissipation to Eq. (\ref{GP}), which is given by
\begin{eqnarray}
\bigl( i-\gamma _{\rm s}(\bm{k}) \bigr) \hbar \frac{\partial}{\partial t}  \tilde{\psi} _{m} (\bm{k})  &=&  \frac{\hbar ^2 k^{2} }{2M} \tilde{\psi} _{m}(\bm{k})  \nonumber \\
&-& \mu _{\rm s} \tilde{\psi} _{m}(\bm{k}) + h_{m}(\bm{k}),
\label{dGPk1}
\end{eqnarray}
\begin{eqnarray}
h(\bm{k}) = \mathcal{F} \large{[} c_{0} \rho (\bm{r}) \psi _{m} (\bm{r}) + c_{1}  \bm{F}(\bm{r}) \cdot {\hat{\bm{F}} } _{mn} \psi _{n} (\bm{r}) \large{]},
\label{dGPk2}
\end{eqnarray}
where $\tilde{\psi} _{m}(\bm{k})$ and $\gamma _{\rm s}(\bm{k}) $ is defined by $\mathcal{F} \large{[} \psi _{m} (\bm{r}) \large{]}$ and $\gamma _{0} \theta (k-k_{\rm s})$.
In the previous study, the chemical potential has a time dependence in order to conserve the total particle number.
However, in our calculation, we do not adjust the chemical potential $ \mu _{\rm s}$, because the total particle number hardly decreases in SST.

\subsection{Numerical parameters and the initial state}
We describe the parameters and the initial state in our numerical calculation of Eqs. (\ref{dGPk1}) and (\ref{dGPk2}).
All our numerical results are obtained in a two-dimensional system, whose size is
$256\xi _{\rho} \times 256\xi _{\rho}$.
To generate SST, we prepare the dynamically unstable state as the initial state, which is the counterflow state \cite{FT12a}. 
This state is given by
\begin{equation}
\begin{pmatrix}
\psi _{1} \\
\psi _{0} \\
\psi _{-1}
\end{pmatrix}
= \sqrt{\frac{\rho _{0}}{2}}
\begin{pmatrix}
{\rm{exp}}(i\frac{MV_{\rm R}}{2 \hbar } x ) \\
0 \\
{\rm{exp}}(-i\frac{MV_{\rm R}}{2 \hbar} x)
\end{pmatrix}
,
\end{equation}
with a relative velocity $V_{\rm R}$.
In our numerical calculation, we set $V_{\rm R}/C_{\rm d} = 24\pi \xi _{\rho}/L \sim 0.294$. 
The dependence of numerical result on $V_{\rm R}$ is discussed in Sec. VI A. 
The usual experimental ratio $| c_{0} / c_{1} |$ for $^{87} {\rm Rb}$ is about $200$ \cite{Rb}, 
but, in our numerical calculation, we set $| c_{0} / c_{1} | = 20$ with a positive $c_{0}$ and negative $c_{1}$ to grow the dynamical instability quickly. We use the strength of the dissipation $\gamma _{0} = 0.03$ and the chemical potential $\mu _{s} = c_{0}\rho _{0}$.   
In the initial state, we add a small white noise contribution to cause the counterflow instability.

%As shown in the following, in this SST, the energy dissipates, so that this turbulence is the decaying turbulence.

\section{Numerical results} 

\subsection{Spectrum of superflow kinetic and spin-dependent interaction energy}

Figures \ref{fig1}(a) and 1(b) show the distribution of the $x$ components of the velocity and spin fields in SST at $t /\tau = 700$
with $\tau = \hbar /c_{0}\rho _{0}$. Through the counterflow instability, these two fields are disturbed.
The $z$ components of the rotation for these two fields are shown in Figs. \ref{fig1}(c) and 1(d), where the rotations are also disturbed 
\cite{vorticity}.
In a spin-1 spinor BEC, the superfluid velocity is related to the spin field through the Mermin-Ho relation,
so that the vortical field can be continuous, which is much different from the one-component BEC \cite{KU,stamper,YU12}.
Actually, as seen in Fig. \ref{fig1}(c), the vortical field $ [{\rm rot} \bm{v}]_{z} $ has a smooth spatial dependence.

The time development of the spectrum of the spin-dependent interaction and the superflow kinetic energy is shown in Fig. \ref{fig2}.
In the early stage of the instability, as shown in Fig. \ref{fig2}(a), these spectra have a peak corresponding to the most unstable wave number 
$k_{\rm in} \xi _{\rho} \sim 0.2 $ for the counterflow instability, which was theoretically obtained by using the Bogoliubov-de Gennes equations for this 
initial state \cite{FT12a}. 
This wave number $k_{\rm in}$ is the energy injection scale.
As time progresses, the energy is transferred from low to high wave number, as seen in Fig. \ref{fig2}(b).
After time $t/\tau = 500$, the spectra of the velocity and the spin-dependent interaction energy exhibit the $-5/3$ and $-7/3$ power laws
in Figs. \ref{fig2}(c) and (d). In Fig. \ref{fig2} (d), the range $0.1 \alt k \xi _{\rho} \alt 0.6$ of these scaling laws is not so wide, 
but the scaling behavior is consistent with our consideration in Sec. III. Therefore, we conclude that our numerical result confirms 
the -5/3 and $-7/3$ power laws. 

We comment on the scaling range. This narrow scaling range may come from how to excite to the initial state. 
As shown in Fig. \ref{fig2}(a), the energy is injected at the wave number 
$k_{\rm in}  \xi _{\rho} \sim 0.2$, which roughly determines the lower limit of the scaling range. 
Therefore, if we use another method with lower $k_{\rm in}$ for generating SST, the scaling range may be wider.  

In our calculation, the $-5/3$ power law applies in the range $ 500< t/\tau <700 $.
For $ t/\tau>800 $, we confirm that the spectrum of the kinetic energy begins to deviate from the $-5/3$ power law as shown in 
Fig. 3; this is caused by the shortage of  energy in the low-wave-number region and the energy accumulation near $k_{\rm s}$.
On the other hand, the $-7/3$ power law in the spectrum of spin-dependent interaction energy sustains even for $ t/\tau>800 $. 
As time sufficiently passes, the spectrum in low wave number region grows, and 
this spectrum has a configuration similar to that in our previous study \cite{FT12a}.
This growth is discussed in Sec. VI. B. 

Finally, we comment on what happens if the energy dissipation is absent. 
We perform numerical calculation without the dissipation, which shows two following behaviors: 
(i) The spectrum of superflow kinetic energy still shows the $-5/3$ power law, 
but the time period sustaining this power law becomes shorter, and (ii) the fluctuation of the spectrum is larger. 
These similar behaviors were observed in the previous study for QT of one-component BEC too \cite{KT05}. 

\begin{figure} [t]
\begin{center}
\includegraphics[width=85mm]{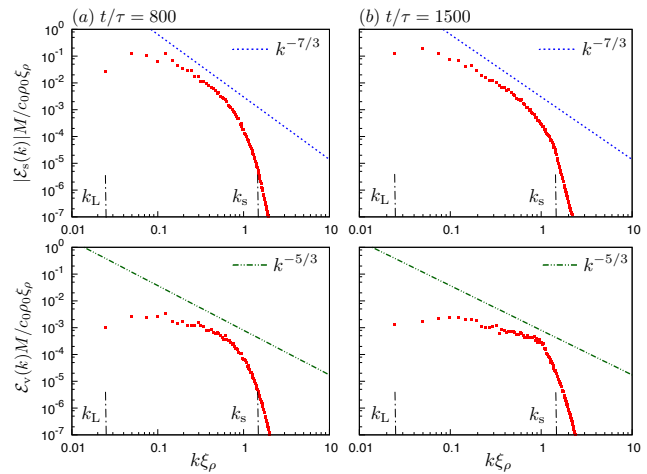}
\caption{(Color online)  Time dependence of the spectrum of spin-dependent interaction (upper) and  superflow kinetic (lower) energy at
$t/\tau =$ (a)  800 and (b) 1500. The dotted lines in the upper and lower graphs are proportional to $k^{-7/3}$ and $k^{-5/3}$, respectively.}
\label{fig3}
\end{center}
\end{figure}

\subsection{Decomposition of kinetic energy}

\begin{figure} [t]
\begin{center}
\includegraphics[width=85mm]{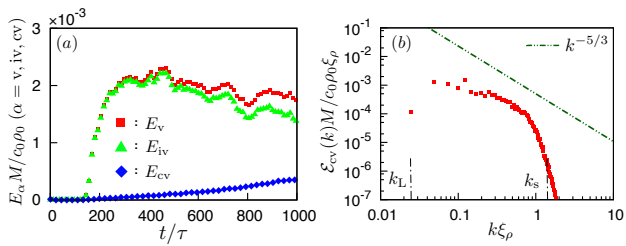}
\caption{(Color online)  (a) Time dependence of $E_{\alpha}$ ($\alpha = {\rm  v, iv, cv}$) and (b) spectrum for compressible
superflow kinetic energy at $t= 700 \tau$. The dotted line in (b) is proportional to $k^{-5/3}$.}
\label{fig4}
\end{center}
\end{figure}

We now consider what structure of velocity field leads to the Kolmogorov $-5/3$ power law in SST.
In QT, the vortical velocity field seems to be important for the Kolmogorov spectrum.
Then, to investigate the vortical flow of $\bm{A} _{\rm v}=\sqrt{\rho} \bm{v}$ in SST, we  decompose
the vector $\bm{A}_{\rm v}$ into  incompressible $ \bm{A}_{\rm iv}$ and compressible $ \bm{A}_{\rm cv}$ parts \cite{Nore}.
The Helmholtz theorem leads to $\bm{A}_{\rm v} = \bm{A}_{\rm iv} + \bm{A}_{\rm cv}$, where the relations ${\rm div} \bm{A}_{\rm iv} = 0$ and ${\rm rot} \bm{A}_{\rm cv} = 0$ are satisfied.
Thus, the superflow kinetic energy per unit mass is expressed 
by $E_{\rm v} = E_{\rm iv} + E_{\rm cv}$ with $E_{\alpha} = \int \bm{A}^{2}_{\alpha} d\bm{r}/2N$ ($\alpha = {\rm v, iv, cv}$).
Using the Fourier series $\bm{A}_{\alpha }(\bm{r}) = \sum \tilde {\bm{A}}_{\alpha }(\bm{k}) e^{i \bm{k} \cdot \bm{r}}$ ($\alpha = {\rm v, iv, cv}$),
we can define the spectra for each kinetic energy per unit mass as
\begin{eqnarray}
\mathcal{E}_{\alpha }(k) = \frac{1 }{2 \rho _{0} \Delta k}  \sum _{k<|\bm{k}_{1}|<k+\Delta k}    |\tilde{\bm{A}}_{\alpha }(\bm{k}_{1})|^{2}.
\label{spectrum_iv}
\end{eqnarray}
We calculate the time dependence of $E_{\alpha}$ in Fig. \ref{fig4}(a); one can see  that the incompressible superflow kinetic energy
is much larger than the compressible one.
This can be caused by (i) the condition $|c_{0}/c_{1}| \gg 1,$ under which the total density is hard to disturb, and
(ii) the dissipation, which prevents the total density modulation from accumulating.
Figure \ref{fig4}(b) shows the spectrum $\mathcal{E}_{\rm cv}$ of the compressible kinetic energy
at $t/\tau = 700$, which deviates from the $-5/3$ power law in comparison with $\mathcal{E}_{\rm v}$ in Fig. \ref{fig2}(d).
Therefore, the vortical structure of $\bm{A}_{\rm v}$ is significant for the Kolmogorov spectrum in SST, which is similar to the situation of QT.

\subsection{-5/3 power law generated by the nonlinear spin term }
We numerically confirm that the $-5/3$ power law in SST is different from that in CT. In Sec. III, we point out that the $-5/3$ power law in SST 
originates from the nonlinear spin term of Eq. (\ref{velocity_motion}).
The magnitude of the inertial and nonlinear spin terms is dependent on the scale, so that we perform the Fourier transform
for these two terms, comparing the magnitude of the Fourier components in order to confirm the argument of Sec. III. 
Specifically, we numerically calculate the following quantities:
\begin{eqnarray}
B_{\mu} (k) =  \sum _{k<|\bm{k}_{1}|<k+\Delta k}  | b_{\mu} (\bm{k}_{1}) | ,
\end{eqnarray}
\begin{eqnarray}
C_{\mu} (k) = \sum _{k<|\bm{k}_{1}|<k+\Delta k}  | c_{\mu} (\bm{k}_{1}) |,
\end{eqnarray}
where $b_{\mu} (\bm{k})$ and $c_{\mu} (\bm{k})$ are defined by 
\begin{eqnarray}
b_{\mu} (\bm{k}) =  \mathcal{F}[ v_{\nu}\nabla _{\nu}v_{\mu}] ,
\end{eqnarray}
\begin{eqnarray}
c_{\mu} (\bm{k}) =  \mathcal{F} \biggl [ \frac{\hbar^{2}}{4M^{2}} \nabla _{\nu}  \biggl\{ (\nabla_{\mu}f_{\lambda})(\nabla_{\nu}f_{\lambda})
 -f_{\lambda}(\nabla_{\mu} \nabla_{\nu} f_{\lambda})  \biggr\} \biggl ].
\end{eqnarray}

Figure 5 shows the wave number dependence of $B_{x}(k)/C_{x}(k)$ and $B_{y}(k)/C_{y}(k)$.  One sees that the nonlinear spin term $C_{\mu}(k)$ 
is larger than the inertial term $B_{\mu}(k)$, and these ratios are about $0.3 \sim 0.07$ in the scaling range 
$0.1 \alt k \xi _{\rho} \alt 0.6$ of Fig. 2 where the $-5/3$ power law appears. 
Thus we can consider that the Kolmogorov spectrum in SST is generated by the nonlinear spin term.

\begin{figure} [t]
\begin{center}
\includegraphics[width=85mm]{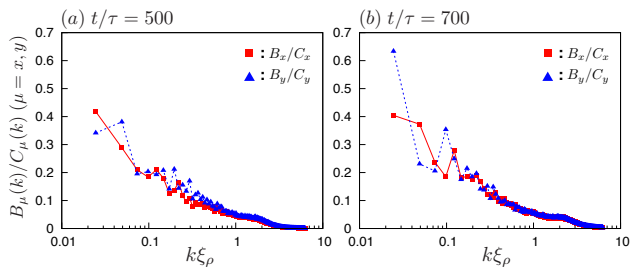}
\caption{(Color online) The ratios $B_{\mu}(k)/C_{\mu}(k)$ ($\mu = x, y$) at $t/\tau =$ (a) 500 and (b) 700. }
\label{fig5}
\end{center}
\end{figure}

\section{Discussions}
\subsection{Dependence of the spectra on $V_{\rm R}$}
We now discuss the dependence of the spectra of the superflow kinetic and spin-dependent interaction energy on $V_{\rm R}$.

When the relative velocity $V_{\rm R}$ is large, the period sustaining the Kolmogorov spectrum in the superflow kinetic energy 
is numerically confirmed to become short. This can be caused by the following two reasons. 
One is (i) modulation of total density, and the other is (ii) high energy injection wave number. 
As for (i), when the relative velocity is large, the total density is easy to fluctuate, which leads to the growth of the compressible velocity field. 
Actually, we numerically confirm the increase of the compressible kinetic energy. 
In our derivation of the power law, the uniformity of the total density is assumed, so that 
this increase shortens the time period sustaining the Kolmogorov spectrum. 
As for (ii), the most unstable wave number becomes high when the relative velocity is large \cite{FT12a}, 
which leads to the high injection wave number $k_{\rm in}$. 
As discussed in Sec. V. A, this generates the narrow scaling range, which disturbs the appearance of the power law.
Because of the above reasons, in our calculation, we set $V_{\rm R}/C_{\rm d} = 24\pi \xi _{\rho}/L \sim 0.294$. 

On the other hand, in the spectrum of spin-dependent interaction energy, the $-7/3$ power law can appear 
even under the condition of large relative velocity. This was confirmed in our previous study \cite{FT12a}.
At present, we do not understood why this spectrum can exhibit the $-7/3$ power law independently of the relative velocity. 

\subsection{Possibility of inverse energy cascade in SST}
In this paper, we perform the two-dimensional numerical calculation for SST, so that the possibility of inverse cascade in SST is discussed.  

As briefly described in Sec. III, in three-dimensional CT, the kinetic energy is 
transported from low wave numbers to high ones, which is called direct energy cascade.
However, in two-dimensional CT, the energy inverse cascade occurs, where the kinetic energy is 
transported from high wave numbers from low ones. This inverse cascade is caused by the conservation of ensthropy 
as well as that of kinetic energy in two-dimensional fluid system \cite{Davidson,Frisch}.  

At present, we consider that the inverse energy cascade for superflow kinetic energy does not occur in the two-dimensional SST because 
the enstrophy in ferromagnetic spin-1 spinor BEC is not conserved. Actually, our numerical calculation 
does not exhibit the sign of the energy inverse cascade in the superflow kinetic energy. 

On the other hand, as for the spin-dependent interaction energy, 
the inverse energy cascade may occur because our previous study and Fig. \ref{fig3} seem to exhibit the growth of 
spectrum of spin-dependent interaction energy in low wave number region as the time sufficiently passes.
This growth may be caused by the ferromagnetic interaction since it tends to align the spin density vector and make the spin domain.   
However, as shown in Fig. \ref{fig2}, the spectrum apparently seems to exhibit the sign of direct energy cascade. 
Thus, at the moment, we does not sufficiently understand the direction of energy cascade for the spin-dependent interaction energy. 

\section{Conclusion} 
We theoretically and numerically studied SST in a spin-1 ferromagnetic spinor BEC at zero temperature by using the spin-1 spinor GP
equations, finding that both the $-5/3$ and $-7/3$ power laws appear in the spectrum of the superflow kinetic and the spin-dependent interaction energy.
First, we discussed the possibility for the Kolmogorov spectrum in SST, pointing out that this spectrum can be generated by the nonlinear
spin term. Second, we showed the numerical results of the spin-1 spinor GP equation with the phenomenological small-scale energy dissipation, whereby SST in the two-dimensional system was obtained by the counterflow instability.
Our numerical results indicated that both the $-5/3$ and $-7/3$ power laws appeared.
Furthermore, we estimated the magnitude of the inertial term
and the nonlinear spin  term of Eq. (\ref{velocity_motion}) in the wave number space, 
numerically confirming that the Kolmogorov spectrum in SST can be generated by the latter term.

\section*{ACKNOWLEDGMENT}
K. F. was supported by a Grant-in-Aid for JSPS Fellows Grant Number 262524. 
M. T. was supported by JSPS KAKENHI Grant Number 26400366 and MEXT KAKENHI "Fluctuation $\&$ Structure" Grant Number 26103526.


\begin{thebibliography}{99}
\bibitem{oe} As shown in Fig. 1, the velocity is smaller than $C_{\rm d}$, whose order is about $0.1 C_{\rm d}$.
The magnitude $|\bm{f}|$ of the spin vector is of the order of unity because of the ferromagnetic interaction. Then, by using the nondimensional form of Eq. (\ref{velocity_motion}), we can estimate the order of each term.
\bibitem{vorticity} In contrast to a one-component BEC, the superfluid velocity and the vorticity in Fig. (\ref{fig1}) do not diverge. 
This is because the vortex core is filled with other components. 
However, if the instability is much stronger, the velocity may diverge since the other components may escape from the vortex core.
\end{thebibliography}
\end{document}